\documentclass[final,3p,times,twocolumn]{elsarticle}
\usepackage{graphicx}% eps fig, package for more complicated commands
\usepackage{amssymb} % various useful mathematical symbols
\usepackage{amsmath} % extended theorem environments
\usepackage{ulem}

\usepackage{lineno}
\usepackage{float} % improved interface for floating objects
\usepackage{subfig} % supersedes the subfigure package
\usepackage{ifpdf}
\ifpdf
\usepackage[pdftex]{hyperref}
\else
\usepackage[hypertex]{hyperref}
\fi

\hypersetup{
  pdftitle={},%
  pdfauthor={},%
  pdfsubject={},%
  pdfkeywords={},%
  pdfstartview={},%
  bookmarksopen=true, breaklinks=true, debug=true, %
  colorlinks=true, linkcolor=red, citecolor=blue, urlcolor=blue
}

\journal{Elsevier; Published in Physics Letters B {\bf 695} (2011) 149-156}

\newcommand{\beq}{\begin{eqnarray}}
\newcommand{\eeq}{\end{eqnarray}}
\newcommand{\be}{\begin{eqnarray*}}
\newcommand{\ee}{\end{eqnarray*}}
\newcommand{\nn}{\nonumber}

\newcommand{\ep}{\varepsilon}

\newcommand{\eqs}[1]{\begin{equation} \begin{split} #1\end{split} \end{equation} }

\newcommand{\ie}{{\it i.e.}}
\newcommand{\eg}{{\it e.g.}}
\newcommand{\etal}{{\it et al.}}

\newcommand{\Br}{{\rm Br}}

\newcommand{\ce}[1]{Eq.~\eqref{#1}}

\newcommand{\cf}[1]{{Fig.~\ref{#1}}}

\newcommand{\Q}{{\cal Q}}

\def\lsim{\raise0.3ex\hbox{$<$\kern-0.75em\raise-1.1ex\hbox{$\sim$}}}
\def\gsim{\raise0.3ex\hbox{$>$\kern-0.75em\raise-1.1ex\hbox{$\sim$}}}

\def\pp   {$pp$}

\def\sqrtsNN {\mbox{$\sqrt{s_{NN}}$}}

\def\jpsi    {\mbox{$J/\psi$}}

\def\beq     {\begin{equation}}
\def\eeq     {\end{equation}}

\long\def\symbolfootnote[#1]#2{\begingroup%
  \def\thefootnote{\fnsymbol{footnote}}\footnote[#1]{#2}\endgroup}

\begin{document}

%%%%%%%%%%%%%%%%%%%%%%%%%%%%%%%%%%%%%%%%%%%%
%% FRONTMATTER
%%%%%%%%%%%%%%%%%%%%%%%%%%%%%%%%%%%%%%%%%%%%

\begin{frontmatter}

\title{QCD corrections to \jpsi\ polarisation in $pp$ collisions
at RHIC}

\author{J.P. Lansberg}

\address{IPNO, Universit\'e Paris-Sud 11, CNRS/IN2P3, 91406 Orsay, France$^\dagger$\\and \\
Centre de Physique Th\'eorique, \'Ecole Polytechnique, CNRS,   91128 Palaiseau, France}

\begin{abstract}
\small
We update the study of the polarisation of  $J/\psi$  produced in proton-proton collisions at  
RHIC at $\sqrt{s}=200$ GeV using the QCD-based Colour-Singlet Model (CSM), including next-to-leading order partonic matrix elements
from gluon and light quark fusion and leading-order contributions from charm-quark initiated processes.  
To do so, we also evaluate the corresponding cross section 
differential in $P_T$ which agrees qualitatively with the measurements of PHENIX in the central and forward regions 
at low $P_T$ -- for instance below 2 GeV --, while emphasising the need
for Initial State Radiation (ISR) resummation. At mid $P_T$, we also compare the measurements from PHENIX and STAR 
with the same evaluation complemented with the dominant $\alpha_S^5$ contributions (NNLO$^\star$). We find a reasonable agreement
 with the data. Regarding the polarisation, as shown for previous studies at larger $\sqrt{s}$ and $P_T$,
the polarisation pattern from gluon and light quark fusion in the helicity frame is 
drastically modified at NLO and is shown to be increasingly longitudinal. 
The yield from charm-gluon fusion is found to be slightly transversally polarised. Combining both these contributions 
with a data-driven range for the polarisation
of $J/\psi$ from $\chi_c$, we eventually provide an evaluation of the polarisation of the prompt $J/\psi$ yield which is in a 
good agreement with the experimental data from PHENIX both in the central and
forward regions.
\end{abstract}


\end{frontmatter}

%%%%%%%%%%%%%%%%%%%%%%%%%%%%%%%%%%%%%%%%%%%%
%% MAINMATTER
%%%%%%%%%%%%%%%%%%%%%%%%%%%%%%%%%%%%%%%%%%%%

\section{Introduction}
\label{sec:intro}

\renewcommand{\thefootnote}{\fnsymbol{footnote}}
\footnotetext[2]{Permanent address}
\renewcommand{\thefootnote}{\arabic{footnote}}

Until recently, the numerous puzzles in the prediction of quarkonium-production 
rates at hadron colliders were attributed to 
non-perturbative effects associated with channels in which the heavy
quark and antiquark are produced in a colour-octet state~\cite{Lansberg:2006dh,Brambilla:2004wf,Kramer:2001hh,Lansberg:2008zm}.
It is now widely accepted that $\alpha^4_s$ and $\alpha^5_s$ corrections to the CSM~\cite{CSM_hadron} 
are fundamental for
understanding the $P_T$ spectrum of $J/\psi$ and $\Upsilon$ produced in
high-energy hadron  collisions~\cite{Campbell:2007ws,Artoisenet:2007xi,Gong:2008sn,Gong:2008hk,Artoisenet:2008fc,Artoisenet:2008zza,Lansberg:2008gk}.
The effect of QCD corrections is also manifest in the polarisation predictions. While the 
$J/\psi$ and $\Upsilon$ (commonly denoted $\Q$ hereafter) produced inclusively or 
in association with a photon are predicted to be 
transversally polarised at LO, it has been recently emphasised that their polarisation at NLO is 
 increasingly longitudinal when $P_T$ gets larger \cite{Gong:2008sn,Artoisenet:2008fc,Li:2008ym,Lansberg:2009db}.

In a recent work~\cite{Brodsky:2009cf}, we have also shown that hard subprocesses 
based on colour singlet $Q \bar Q$ configurations alone are
sufficient to account for the observed magnitude of the $P_T$-integrated
cross section. In particular, the predictions at
LO~\cite{CSM_hadron} (\cf{diagrams} (a)) and NLO~\cite{Campbell:2007ws,Artoisenet:2007xi,Gong:2008sn} 
(\cf{diagrams} (b,c,d)) accuracy
are both compatible with the measurements by the PHENIX collaboration at RHIC~\cite{Adare:2006kf,RHIC2009}
within the present uncertainties. This provided  some indications that the computations are carried in a proper perturbative regime.
This agreement is improved when hard subprocesses involving the 
charm-quark distribution of the colliding protons are taken into consideration. These 
constitute part of the LO ($\alpha_S^3$) rate (\cf{diagrams} (e)) and are
responsible for a significant fraction of the observed yield, as we have 
argued in~\cite{Brodsky:2009cf}.

\begin{figure}[b!]
\vspace*{-1cm}
\centering
\subfloat[]{\includegraphics[scale=.3]{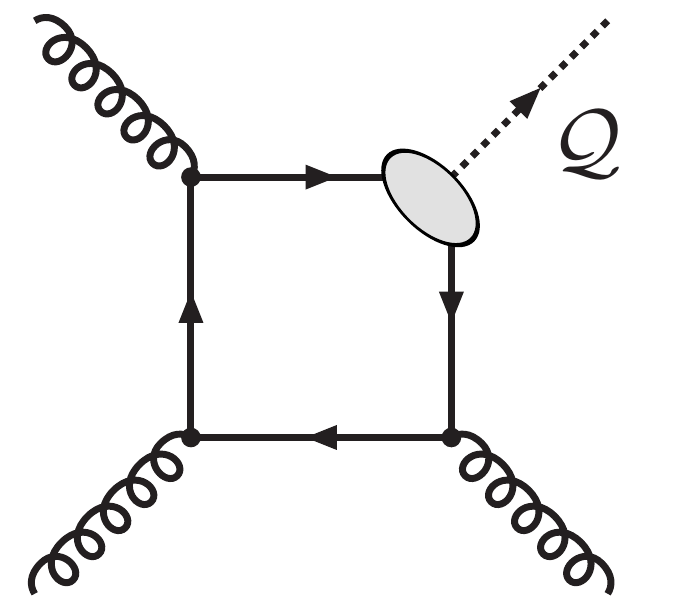}}\hspace*{-.2cm}
\subfloat[]{\includegraphics[scale=.3]{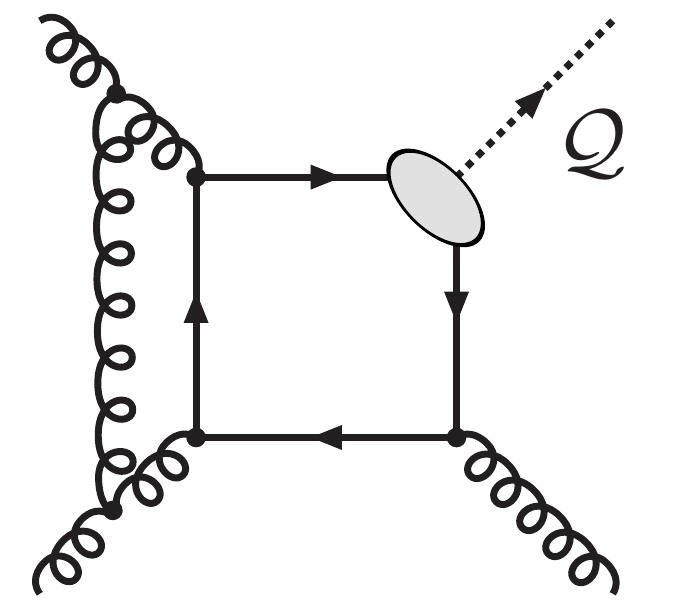}}\hspace*{-.2cm}
\subfloat[]{\includegraphics[scale=.3]{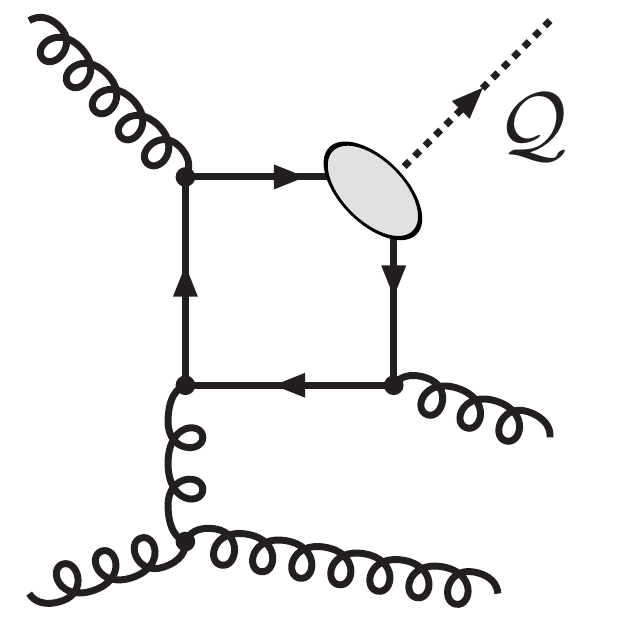}}\hspace*{-.2cm}\\
\subfloat[]{\includegraphics[scale=.3]{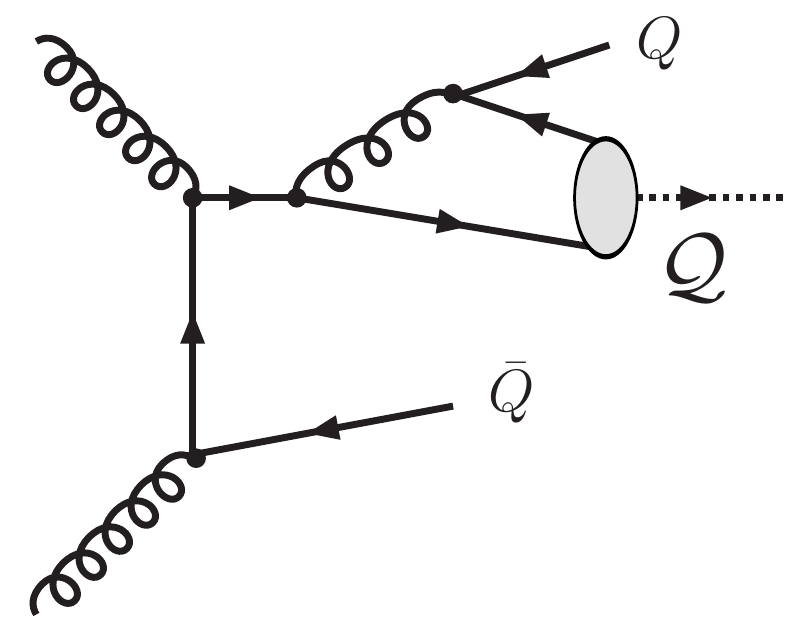}}\hspace*{-.2cm}
\subfloat[]{\includegraphics[scale=.3]{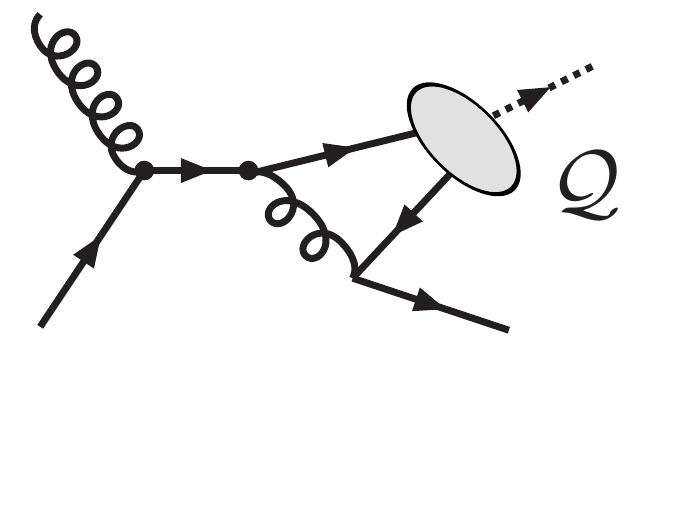}}\hspace*{-.2cm}
\subfloat[]{\includegraphics[scale=.3]{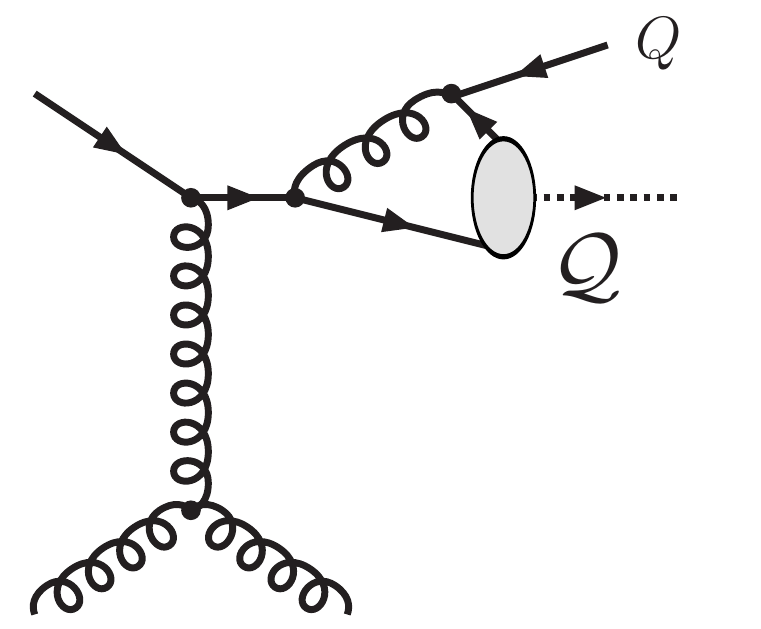}}
\caption{Representative diagrams contributing to $^3S_1$ quarkonium {(denoted $\Q$)} hadroproduction in the CSM
 by gluon fusion at orders $\alpha_S^3$ (a), $\alpha_S^4$ (b,c,d) and initiated
by a charm quark at orders $\alpha_S^3$ (e), $\alpha_S^4$ (f).
The quark and antiquark attached to the ellipsis are taken as on-shell
and their relative velocity $v$ is set to zero.}
\label{diagrams}
\end{figure}

In this Letter, we proceed to the evaluation of the $P_T$ dependence 
of the polarisation of the \jpsi\ produced at RHIC, both in the central and forward rapidity regions. 
In the section 2, we expose the procedure used to compute the yield and the polarisation 
at NLO (up to $\alpha_S^4$) from $gg$ \& $gq$ fusion and from  $cg$ fusion at LO (at $\alpha_S^3$) as well 
as the procedure to obtain a first evaluation of the leading $P_T$ contributions from $gg$ and $gq$ fusion
at $\alpha_S^5$ (NNLO$^\star$). In the section 3, we present our results. First, we show  the yields differential in the rapidity $y$
and $P_T$ along with the polarisation vs $P_T$ from $gg$ \& $gq$ and from $cg$ fusion separately. Then we show our results
 when they are combined. Finally we combine the latter predictions for the direct yield  polarisation 
with an essentially data-driven estimation of the polarisation for $J/\psi$ from $\chi_c$ and we compare 
our results with the PHENIX data in both rapidity ranges. The last section gathers our conclusions.

\section{Cross-section and polarisation evaluation }

\subsection{Cross section}

In the CSM~\cite{CSM_hadron}, the matrix element to create
a $^3S_1$
 quarkonium ${\Q}$ of momentum $P$ and polarisation $\lambda$
 accompanied by
other partons, noted $j$, is the product of the amplitude to create
the corresponding heavy-quark pair, a spin
 projector $N(\lambda| s_1,s_2)$ and
$R(0)$, the radial wave function at the origin in the configuration
space, obtained from the leptonic width~\cite{Amsler:2008zzb}, namely 
\eqs{ \label{CSMderiv3}
{\cal M}&(ab \to {\Q}^\lambda(P)+j)=\!\sum_{s_1,s_2,i,i'}\!\!\frac{N(\lambda| s_1,s_2)}{ \sqrt{m_Q}} \frac{\delta^{ii'}}{\sqrt{N_c}} 
\frac{R(0)}{\sqrt{4 \pi}}\\\times&
{\cal M}(ab \to Q^{s_1}_i \bar Q^{s_2}_{i'}(\mathbf{p}=\mathbf{0}) + j),
}
where $P=p_Q+p_{\bar Q}$, $p=(p_Q-p_{\bar Q})/2$, 
$s_1$,$s_2$ are the heavy-quark spin and $\delta^{ii'}/\sqrt{N_c}$ is the projector onto a colour-singlet state.
In the non-relativistic limit, $N(\lambda| s_1,s_2)$
can be written as 
$ \frac{\ep^\lambda_{\mu} }{2 \sqrt{2} m_Q } \bar{v} (\frac{\mathbf{P}}{2},s_2) \gamma^\mu u (\frac{\mathbf{P}}{2},s_1) \,\, $
where $\ep^\lambda_{\mu}$ is the polarisation vector of the quarkonium. The sum over the quark spin yields to traces
evaluated in a standard way.

In our evaluation, we use the partonic matrix elements from Campbell,
Maltoni and Tramontano~\cite{Campbell:2007ws} to compute the
LO and NLO cross sections from gluon-gluon and light-quark gluon fusion. We guide the reader to~\cite{Campbell:2007ws}
for details concerning the derivation of the cross section at $\alpha_S^4$, the corresponding 
expressions at $\alpha_S^3$ can be found in~\cite{Baier:1983va}. 
In the case of the $cg$ fusion {(at LO)}, we use the framework described in~\cite{Artoisenet:2007qm} based on the
tree-level matrix element generator {\small MADONIA}~\cite{Madonia}. 

Moreover, in order to illustrate the expected impact of NNLO QCD corrections for increasing
$P_T$, we also present the results
when the leading-$P_T$ contributions at $\alpha_S^5$ evaluated along the lines of~\cite{Artoisenet:2008fc} 
are added to the previous contributions. At this order, the last kinematically-enhanced topologies open up. These exhibit
a $P_T^{-4}$ fall off of their differential cross section in $P_T^2$, typical of a single particle exchange in the $t$ channel. 
Going further in the $\alpha_S$ expansion cannot bring any further kinematical enhancement as regards the $P_T$ dependence. 
As a consequence, above $\alpha_S^5$, usual 
expectations for the impact of QCD corrections would then hold and these should be then well taken into account by 
a $K$ factor multiplying the yield at NNLO accuracy, which would be independent of $P_T$ and of a similar magnitude as those of 
other QCD processes. In other words, a further 
cross-section modification between the NNLO and N$^3$LO results by an order of magnitude would not be acceptable at any $P_T$, while 
it is expected to be so between the cross 
sections at LO ($P_T^{-8}$), NLO ($P_T^{-6}$) and NNLO ($P_T^{-4}$) for  large enough $P_T$, simply 
owing to their different $P_T$ scalings.  

The procedure used here to evaluate the leading-$P_T$ NNLO contributions is exactly the same as 
in~\cite{Artoisenet:2008fc}. Namely, the 
real-emission contributions at $\alpha_S^5$ are evaluated using {\small MADONIA} by imposing a lower bound on the invariant-mass of 
any light partons ($s_{ij}$). For the new channels opening up at $\alpha^5_S$ with a leading-$P_T$ behaviour, 
and which specifically interest us, the 
dependence on this cut is to get smaller for large $P_T$ since no collinear or soft divergences can appear there.
For other channels, whose Born contribution is at $\alpha^3_S$ or $\alpha^4_S$, the cut would 
produce logarithms of $s_{ij}/s_{ij}^{\rm min}$, which are not necessarily small. Nevertheless, 
they can be factorised over their corresponding Born contribution, which scales as $P_T^{-8}$ or $P_T^{-6}$, and are thence 
suppressed by at least two powers of $P_T$ with respect of the leading-$P_T$ contributions ($P_T^{-4}$). The sensitivity on 
$s_{ij}^{\rm min}$ is thus expected to come to nothing at large $P_T$. This argument has been checked at  
$\alpha^4_S$ (NLO vs. NLO$^\star$) for 
$\Upsilon$~\cite{Artoisenet:2008fc} and $\psi$~\cite{Lansberg:2008gk} in the inclusive case as well as in association with a photon~\cite{Lansberg:2009db}.

For the parameters entering the cross-section evaluation, we have taken $|R_{J/\psi}(0)|^2=1.01$ GeV$^3$.
We also take Br$(J/\psi \to \ell^+\ell^-)=0.0594$.  
Neglecting relativistic corrections, one has in the CSM, $M_{J/\psi}=2m_c$. 
The uncertainty bands for the resulting predictions at LO and NLO are obtained from the {\it combined}
 variations of the heavy-quark
mass within the ranges $m_c=1.5\pm 0.1$ GeV, the
factorisation $\mu_F$ and the renormalisation $\mu_R$ scales
 chosen\footnote{In principle, the renormalisation scale ambiguity can be removed using the method described
  in \cite{Binger:2006sj}.}
in the couples $((0.75,0.75);(1,1);(1,2);(2,1);(2,2))\times m_T$ with $m^2_T=4m_Q^2+P_T^2$.
The band for the NNLO$^\star$ is obtained for simplicity using a combined variation of $m_c$, $0.5 m_T <\mu_R=\mu_F< 2 m_T$
and $2.25  < s_{ij}^{\rm min}<   9.00$ GeV$^2$.

\subsection{Polarisation}

The  polarisation parameter $\alpha$ (also called $\lambda$) is computed by
analysing the distribution of the polar  angle $\theta$ between the $\ell^+$ direction in the
quarkonium rest frame and the quarkonium direction in the laboratory
frame. This definition of $\theta$ is referred to the analysis of the polarisation 
(or spin-alignement) in the helicity frame\footnote{Other choices are possible and the reader is guided to~\cite{Faccioli:2008dx} for more details.}
.

 The normalised angular distribution $I(\cos \theta)$  reads
\begin{equation}
\label{eq:angulardist}
I(\cos \theta) =
\frac{3}{2(\alpha+3)} (1+\alpha \, \cos^2 \theta)\,,
\end{equation}
from which we can extract $\alpha$  bin by bin in $y$ or $P_T$. 
The interpretation of $\alpha$ is immediate when one relates it to the polarised
cross sections:
\eqs{\label{eq:alph-pol-sigma}
\alpha=\frac{\sigma_T-2\sigma_L}{\sigma_T+2\sigma_L}.
}

\begin{figure}[b!]
\begin{center}
\includegraphics[width=\columnwidth]{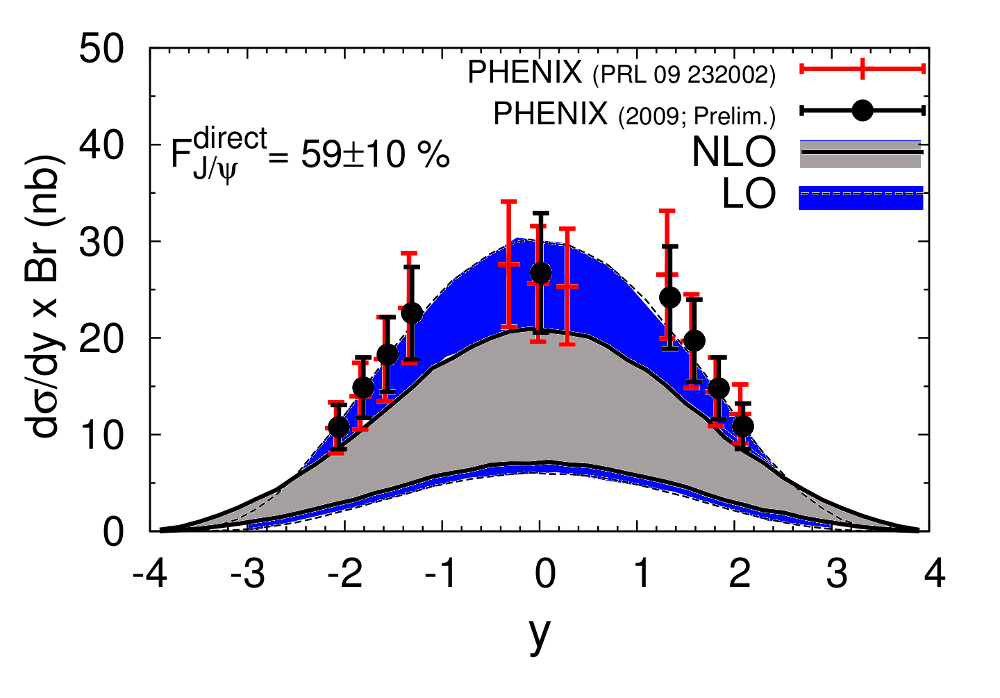}
\end{center}
\caption{$d\sigma/dy \times {\rm Br} $ in \pp\ collisions at $\sqrtsNN=200\mathrm{~GeV}$
at LO and NLO accuracy compared to the PHENIX data~\cite{Adare:2006kf,RHIC2009}. The theoretical-error bands 
for LO and NLO  come from combining the
uncertainties resulting from the choice of $\mu_f$, $\mu_r$, $m_q$. We have used the LO set 
{\small CTEQ6\_L}~\cite{Pumplin:2002vw} for the yield at LO,
the NLO set {\small CTEQ6\_M}~\cite{Pumplin:2002vw} for the yield at NLO.}
\label{fig:dsigdy-gg}
\end{figure}

For $\sigma_T \gg \sigma_L$,
thus for a yield purely transversally polarised, $\alpha\simeq 1$ while for 
$\sigma_L \gg \sigma_T$ (a yield purely longitudinally polarised), $\alpha\simeq -1$.
On the contrary, if no direction is favoured, one expects $\sigma_T = (\sigma_{T_x}+\sigma_{T_y})= 2 \sigma_L$, which corresponds
to  $\alpha\simeq 0$, and thus no $\theta$ dependence. We emphasise here that the relations
\ce{eq:angulardist} and \ce{eq:alph-pol-sigma} do not depend on the definition chosen for 
$\theta$ (the frame definition). Yet, the results obtained do depend on it: a yield transversally polarised in one
frame can be longitudinal in another. In particular, the expressions that one would obtain
for $\sigma_L$ and $\sigma_T$  change from one frame to another.

\section{Results}

\subsection{$gg$ and $gq$ channels}

We first present the results from the $gg$ and $gq$ channels at LO ($\alpha_S^3$) and NLO ($\alpha_S^3$+$\alpha_S^4$) in 
terms of differential cross sections as function of $y$ and $P_T$, which we compare to 
the experimental data for prompt $J/\psi$ multiplied by the expected fraction of 
direct $J/\psi$ ($59\pm 10)\%$ (see~\cite{Brodsky:2009cf}).

\begin{figure*}[!htb]
\begin{center}
\subfloat[central]{\includegraphics[width=\columnwidth]{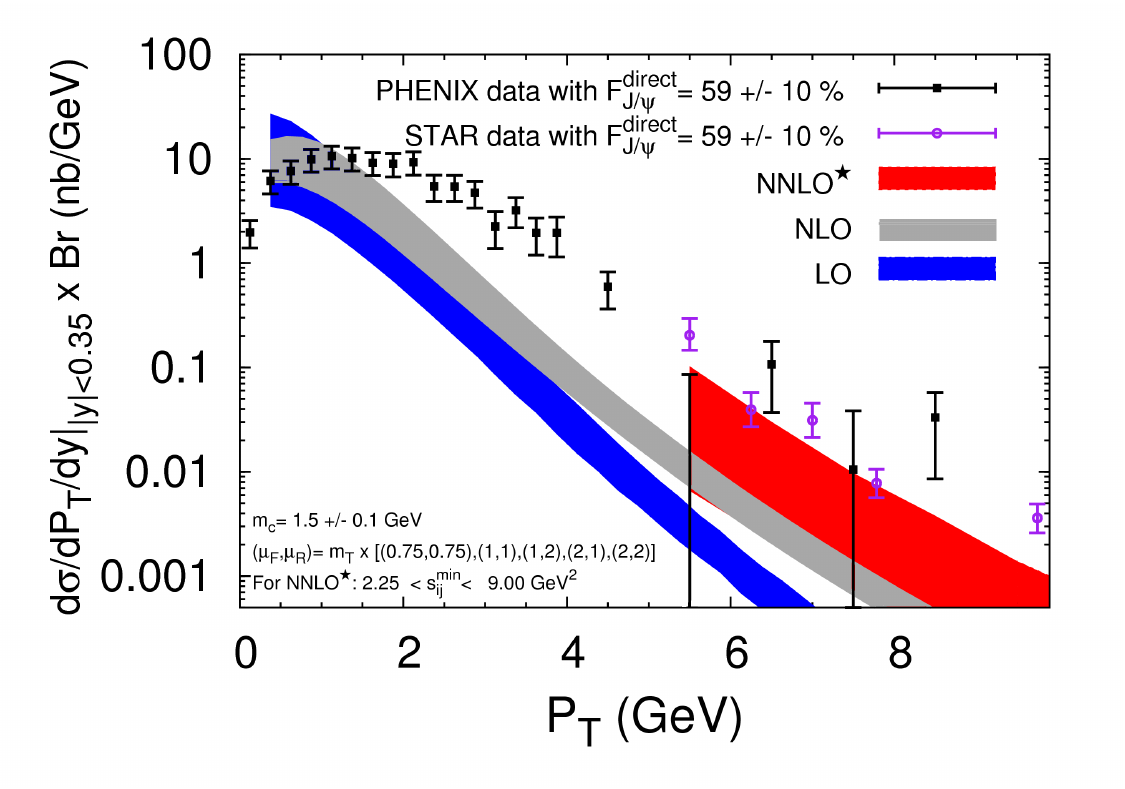}}
\subfloat[forward]{\includegraphics[width=\columnwidth]{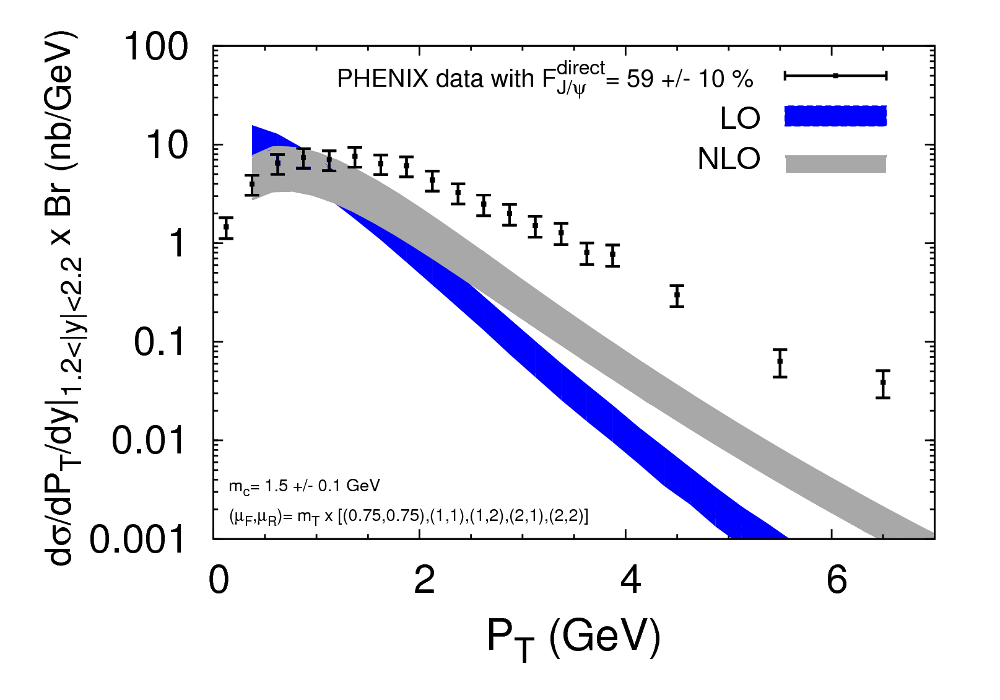}}
\end{center}
\caption{ $d\sigma/dP_T/dy \times {\rm Br}$ from the $gg$ and $gq$ fusion channels in \pp\ collisions at various order in $\alpha_S$
at $\sqrtsNN=200\mathrm{~GeV}$: a) in the central ($|y|<0.35$), b) in the forward ($1.2 < |y|< 2.2$) regions
compared to the PHENIX~\cite{Adare:2006kf} and STAR~\cite{Abelev:2009qa} data. The theoretical-error bands 
come from combining the uncertainties resulting from the choice of $\mu_f$, $\mu_r$, $m_q$, see text.}
\label{fig:dsigdpt-gg}
\end{figure*}

As discussed in~\cite{Brodsky:2009cf}, the experimental data for the $P_T$ integrated cross section
from PHENIX~\cite{Adare:2006kf,RHIC2009} are qualitatively well reproduced by the highest value
of the theoretical bands, both at LO and NLO (see~\cf{fig:dsigdy-gg}). Let us also note that choosing  a lower value than
1.4 GeV for the charm quark mass   -- along the same lines as what has been done in studies of charm
 production at RHIC~\cite{Cacciari:2005rk}-- would lead to 
higher cross sections. However a part of the increase of the cross section could be attributed to the approximation $M_{J/\psi}=2m_c$, 
which would then induce an artificially low bound-state mass in the kinematics. Such an effect is not easily quantifiable 
without a dedicated study. Therefore, we do not use values lower than 1.4 GeV for $m_c$ here.

Before discussing the results of the cross section differential in $P_T$ at LO and NLO, it is important 
to note that we are working in the low $P_T$ region where the perturbative expansion is not always reliable.
While the comparison between the results at LO and NLO integrated in $P_T$ shown on \cf{fig:dsigdy-gg}
gives some good indication that the perturbative expansion works well, it may not be so when one considers
the cross section differential in $P_T$. First, the yield at NLO has shown negative --thence unphysical-- values
in some bins in $P_T$ and $y$. When this occurs, this is always for the lowest $P_T$ bin. This shows
that the virtual corrections at $\alpha_S^4$ (with a negative relative sign) 
are large. This may also be the case for the  virtual 
corrections at $\alpha_S^5$  which are currently unknown and possibly with an opposite sign to the ones at $\alpha_S^4$. This is a known issue which can be partly solved
by resumming Initial State Radiation (ISR). Such a resummation has been carried out for the $\Upsilon$
production at the Tevatron using the Colour Evaporation Model~\cite{Berger:2004cc}. One of the main outcomes
of the latter study is that the $P_T$ dependence of the cross section is significantly affected at 
low $P_T$ (up to roughly $m_{\Upsilon}/2$) by ISR. A similar impact of ISR is expected for the charmonium family,
albeit for a narrower range in $P_T$.

\begin{figure*}[!htb]
\begin{center}
\subfloat[central]{\includegraphics[width=\columnwidth]{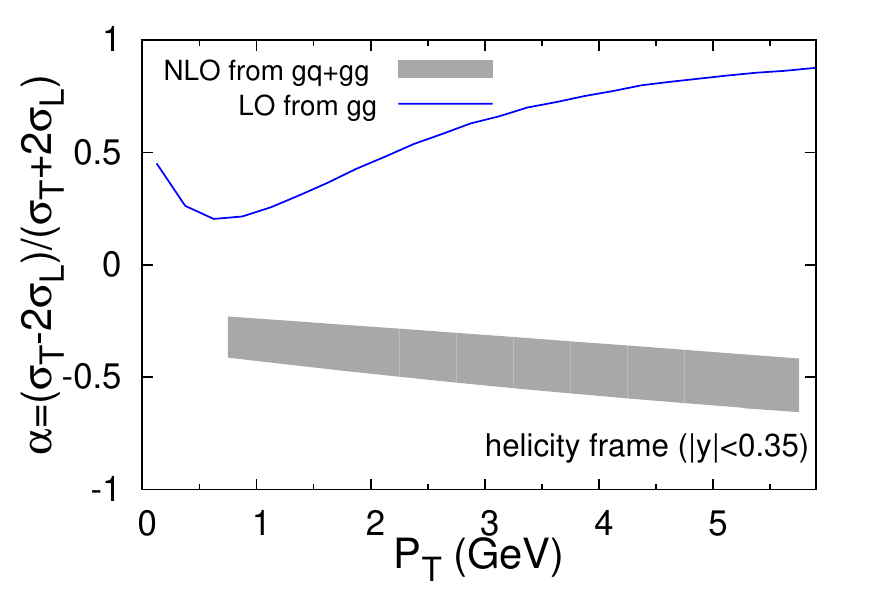}}
\subfloat[forward]{\includegraphics[width=\columnwidth]{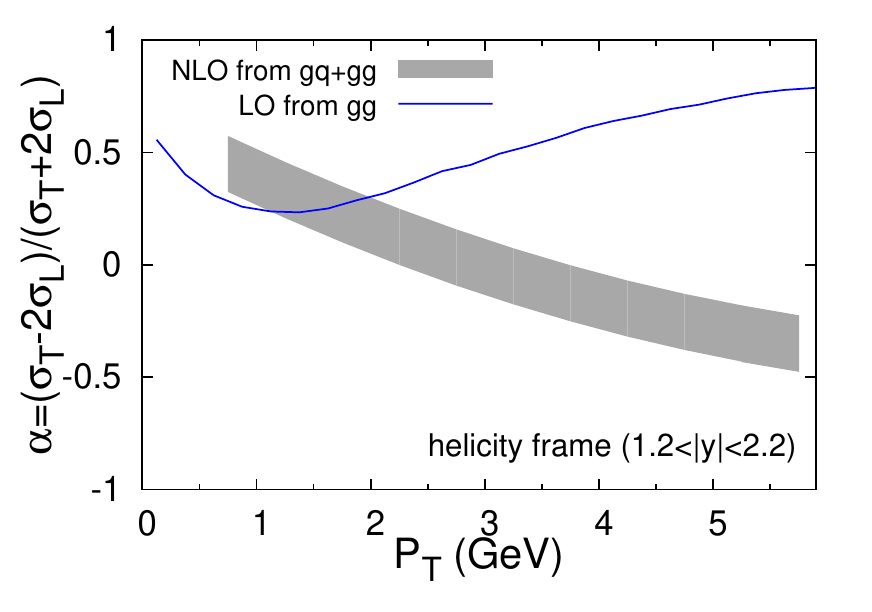}}
\end{center}
\caption{$\alpha(P_T)$ at LO and NLO from $gg$ and $gc$ fusion in \pp\ at $\sqrtsNN=200\mathrm{~GeV}$: 
a) in the central ($|y|<0.35$), b) in the forward ($1.2 < |y|< 2.2$) regions.}
\label{fig:pol_vs_PT}
\end{figure*}

Since we have not carried out this resummation\footnote{Such a resummation is beyond 
the scope of this work and will be the object of future investigations.}, the results that we 
have obtained for $P_T$ below $m_{J/\psi}/2$ for $d\sigma/dP_T$ shown on \cf{fig:dsigdpt-gg} 
(as well as later for the polarisation) are to be taken with a grain of salt.
That being said, one sees on \cf{fig:dsigdpt-gg} that the slope for the yield at NLO is milder than for the LO, 
though too steep to 
reproduce the data from PHENIX~\cite{Adare:2006kf} and STAR~\cite{Abelev:2009qa}. Yet, the general effect of 
ISR is to lower the cross section at very low $P_T$ and to
rise it a little up to $P_T \simeq 1-2 m_c$.

In addition, one expects a significant contribution
from $\alpha_S^5$ contributions at larger $P_T$. A complete evaluation of NNLO corrections is not yet available. 
For now, we can only rely on a study of the leading-$P_T$ NNLO contributions (NNLO$^\star$) as done in~\cite{Artoisenet:2008fc,Lansberg:2008gk}. However, 
for such low values of $P_T$, the latter method is a priori not reliable and we
 could not extend its evaluation (red band) in~\cf{fig:dsigdy-gg} (a)~\footnote{We did not show the NNLO$^\star$ 
band in~\cf{fig:dsigdpt-gg} (b) since the range in $P_T$ is limited: STAR measurements are focused on the central 
region for the time being.} below $P_T=5$ GeV. Overall, the comparison presented allows us to think
that a more complete analysis (through ISR resummation and matching with leading $P_T$ contributions 
for instance) could show that the $P_T$ dependence 
given by the CSM agrees with the experimental measurements in the low and mid $P_T$ regions at RHIC energies.

We now turn to the discussion of the polarisation results vs $P_T$ (\cf{fig:pol_vs_PT}) in the helicity frame. 
As can be seen on \cf{fig:pol_vs_PT}, the complete modification of the polarisation pattern between the LO
and the NLO results observed in previous works~\cite{Gong:2008sn,Artoisenet:2008fc,Li:2008ym,Lansberg:2009db} is confirmed, 
despite the error band due to the usual mass and scale dependence on top of statistical fluctuations (see below). The 
LO transverse yield becomes  increasingly longitudinal at NLO for increasing $P_T$. In the forward region, 
the yield at NLO is transverse at low $P_T$ and becomes longitudinal as soon as  $P_T \geq m_c$.

As discussed previously, the result obtained in the lowest $P_T$ bin (below 1 GeV)
are likely to be subject to higher QCD corrections and the band should not be extrapolated down to $P_T=0$. 
Second, the extraction of 
the yield polarisation at NLO at low $P_T$ is highly demanding in terms of computer time.
Typically, the evaluation of the yield polarisation in a single $P_T$ bin of 500 MeV with statistical fluctuation less than 20 $\%$ 
requires ${\cal O}(10^8)$ numerical evaluations of the integrand. These fluctuations add up to the usual theoretical uncertainties 
in the bands shown on ~\cf{fig:pol_vs_PT}.

\subsection{$cg$ channel}

We first present the results from the $cg$  LO ($\alpha_S^3$) in 
terms of differential cross sections as function of $y$ and $P_T$, which we compare to 
the experimental data for prompt $J/\psi$ multiplied by the expected fraction of 
direct $J/\psi$.

\begin{figure}[!b]
\includegraphics[width=\columnwidth]{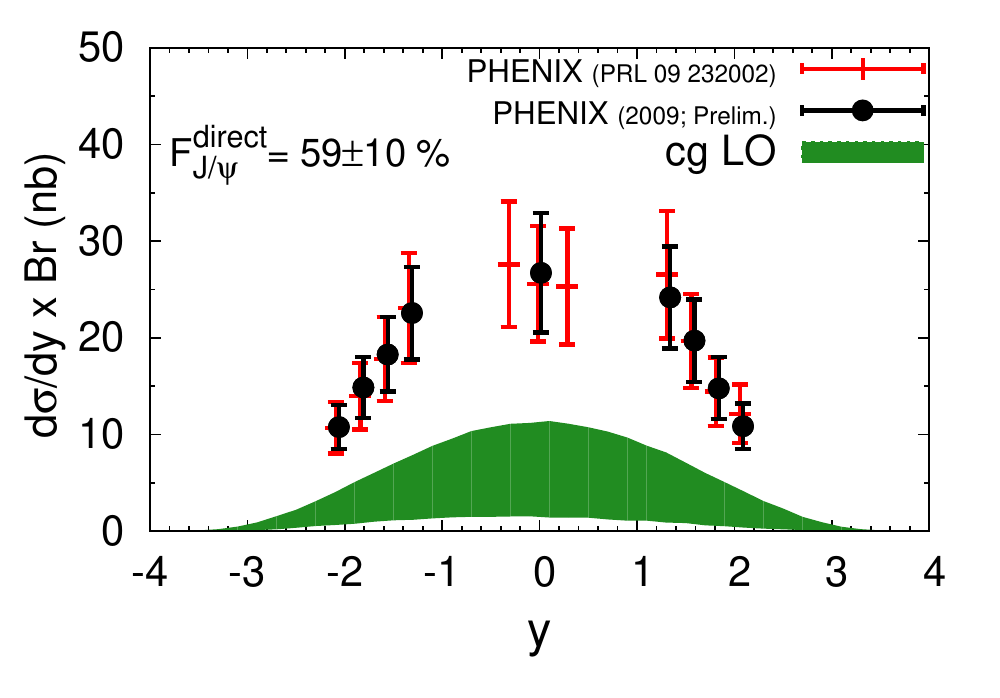}
\caption{$d\sigma/dy \times {\rm Br} $ from $cg$ fusion in \pp\ at $\sqrtsNN=200\mathrm{~GeV}$ using
a sealike charm distribution compared to the PHENIX data~\cite{Adare:2006kf}. The theoretical-error bands 
for LO and NLO  come from combining the
uncertainties resulting from the choice of $\mu_f$, $\mu_r$, $m_q$.}
\label{fig:plot-dsigdy-cg}
\end{figure}

\begin{figure*}[!htb]
\begin{center}
\subfloat[central]{\includegraphics[height=.9\columnwidth,angle=-90]{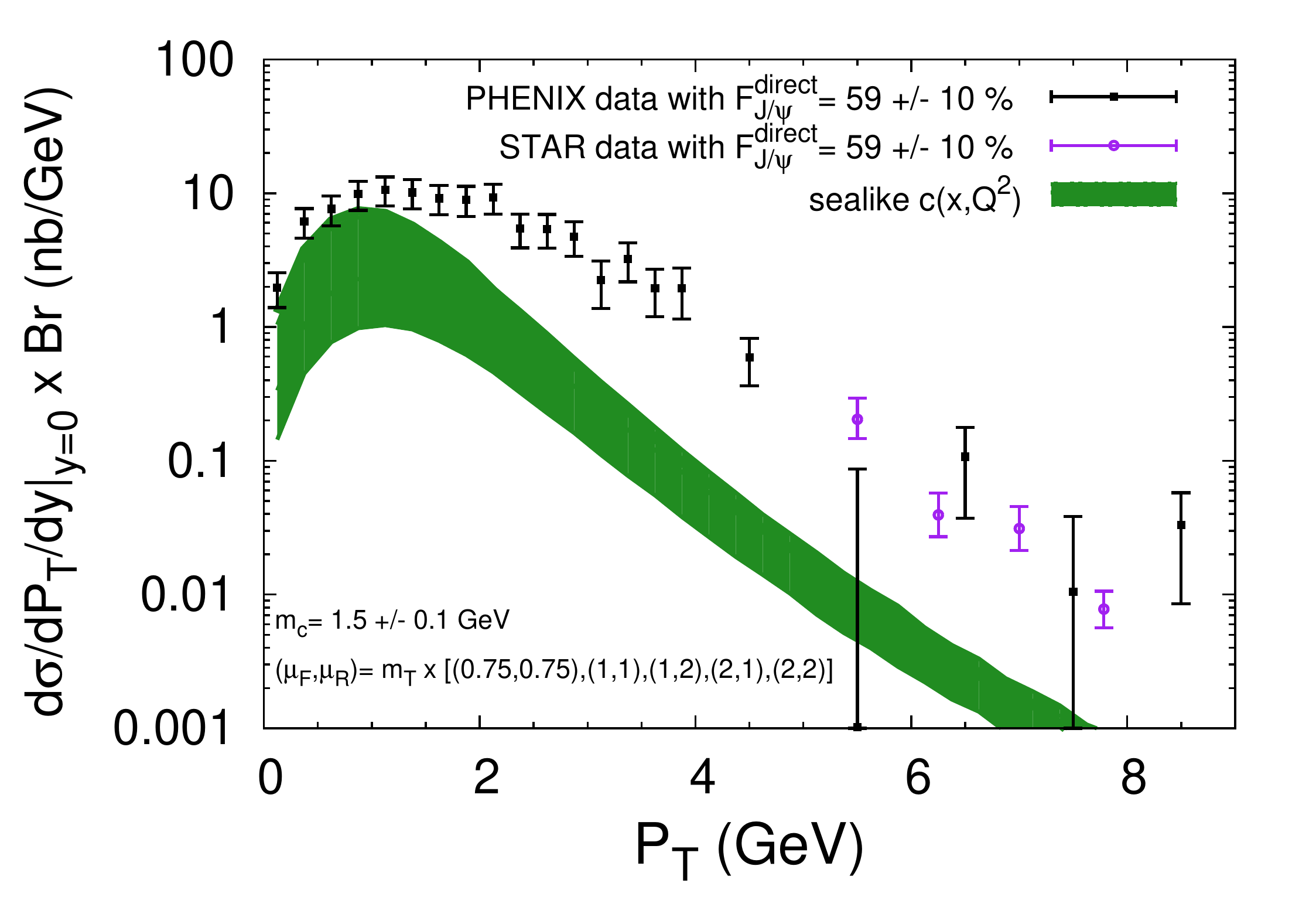}}
\subfloat[forward]{\includegraphics[height=.9\columnwidth,angle=-90]{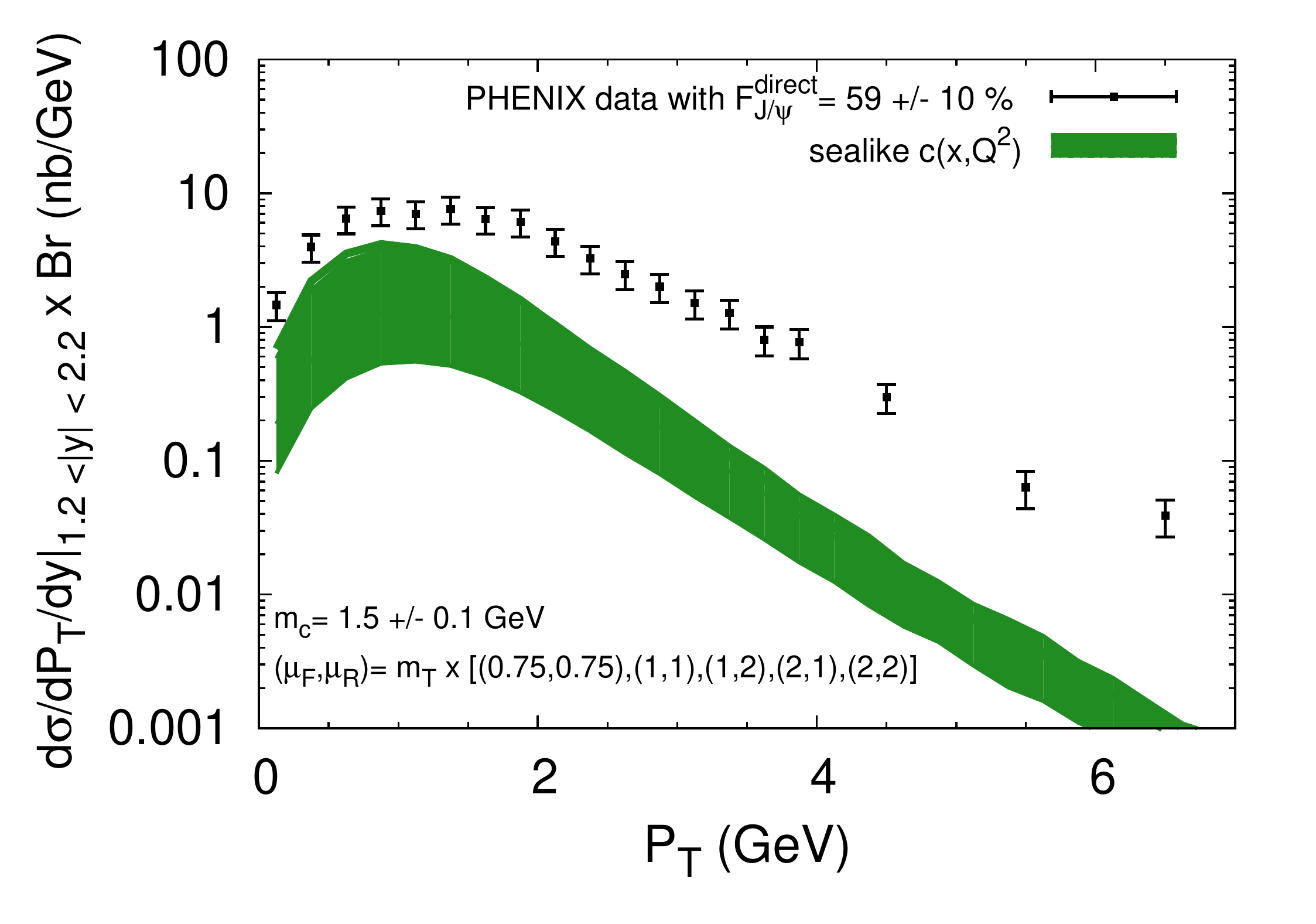}}
\end{center}
\caption{ $d\sigma/dP_T/dy \times {\rm Br}$ from $cg$ fusion in \pp\ at $\sqrtsNN=200\mathrm{~GeV}$ using
a sealike charm distribution in the central and forward rapidity regions compared to the PHENIX~\cite{Adare:2006kf}
and STAR data~\cite{Abelev:2009qa} data. The theoretical-error bands 
come from combining the uncertainties resulting from the choice of $\mu_f$, $\mu_r$, $m_q$.}
\label{fig:plot-dsigdpt-cg}
\end{figure*}

\begin{figure*}[!htb]
\begin{center}
\subfloat[central]{\includegraphics[height=.9\columnwidth,angle=-90]{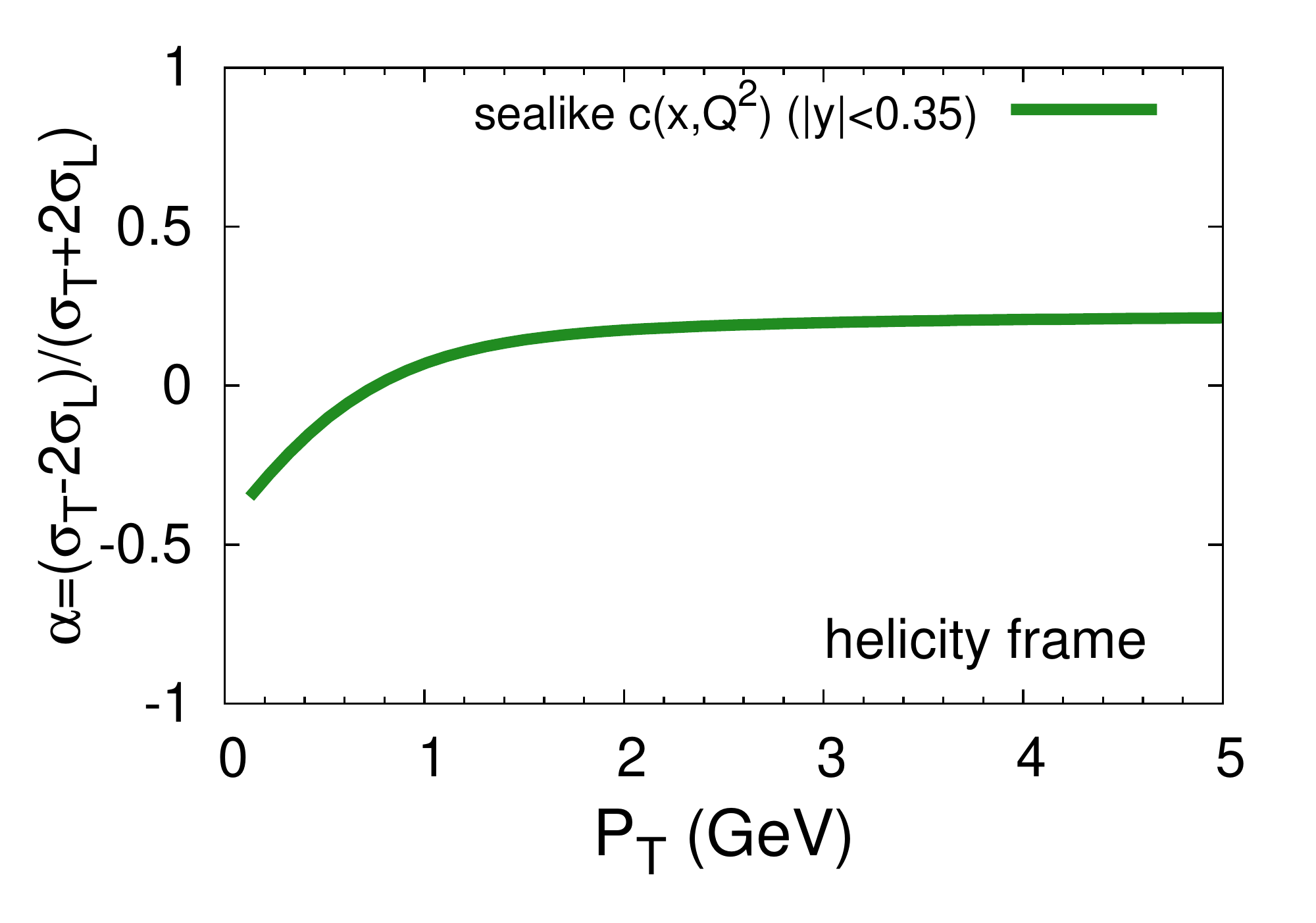}}
\subfloat[forward]{\includegraphics[height=.9\columnwidth,angle=-90]{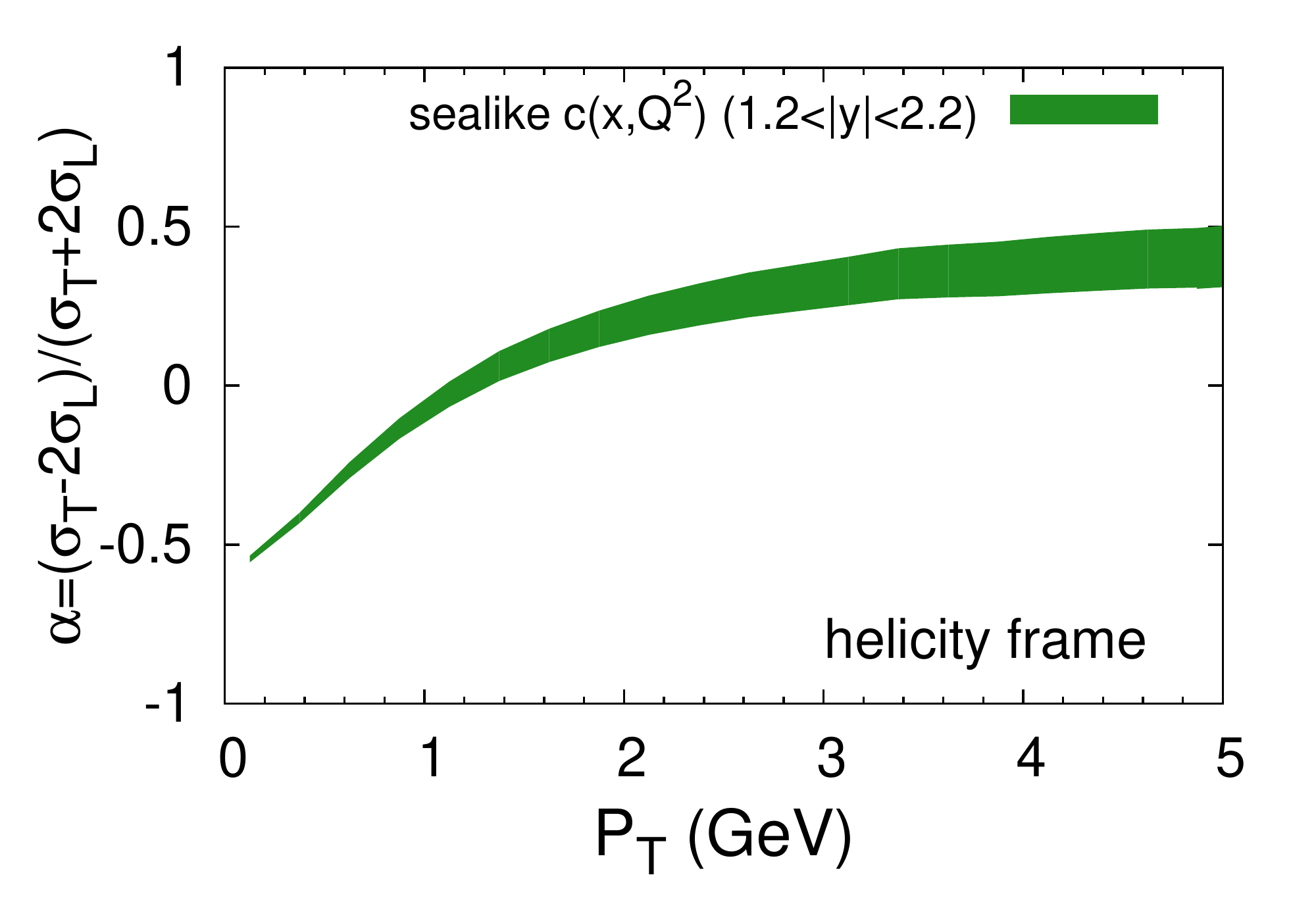}}
\end{center}
\caption{$\alpha(P_T)$ for $J/\psi$ directly produced by $cg$ fusion in \pp\ at $\sqrtsNN=200\mathrm{~GeV}$ using
a sealike charm distribution.}
\label{fig:pol-cg}
\end{figure*}

In~\cite{Brodsky:2009cf}, we have computed the yield integrated over $P_T$ 
from $cg$ channels and used the LO set {\small CTEQ6.5c}~\cite{Pumplin:2007wg}
 based on a recent global PDF fit including Intrinsic Charm (IC). More precisely, we have compared the results for three 
choices for the charm distribution: (i) without IC [$c(x,\mu_0)=0$ ($\mu_0=1.2$ GeV)], (ii) with  BHPS IC~\cite{Brodsky:1980pb}
($\langle x \rangle_{c+\bar c}{\equiv\int^1_0  x [c(x)+\bar c(x)] dx=}2\%$) and (iii) with sealike IC ($\langle x \rangle_{c+\bar c}=2.4\%$). In the following, we shall carry on our study of the polarisation with the sealike IC. The result are qualitatively 
the same with other choice for $c(x)$.

By comparing the data to our results for the contribution from $cg$ fusion shown on~\cf{fig:plot-dsigdy-cg}, one sees that 
it accounts for 5 to 40 $\%$ of the observed yield depending on the usual theoretical uncertainties.
Comparable fractions are obtained for other $c(x)$ as discussed in~\cite{Brodsky:2009cf}.

Regarding the $P_T$ dependence shown in \cf{fig:plot-dsigdpt-cg} for both rapidity regions, one observes that the contribution
is falling too fast compared to the PHENIX data. Nevertheless, the appearance
of NLO contribution such as $cg\to J/\psi cg$ (\eg~\cf{diagrams} (f)) at larger\footnote{These were for instance analysed 
in the fragmentation approximation in~\cite{Qiao:2003pu}.} $P_T$, 
and to a lesser extent the effect of ISR at small $P_T$, is expected to smear the curve out to larger $P_T$.

To what concerns the polarisation pattern (\cf{fig:pol-cg}), in the central region, it is similar to the one computed for 
$J/\psi + c \bar c$ in~\cite{Artoisenet:2007xi}, 
namely nearly unpolarised as soon as $P_T\geq m_c$, although not completely unpolarised
with $\alpha \simeq 0.2$ for mid $P_T$. In the forward region, it becomes somewhat more transversally polarised and
shows a dependence on the mass and scale choices as indicated by the widening band.

\subsection{Polarisation of the direct yield}

\begin{figure*}[htb!]
\begin{center}
\subfloat[central]{\includegraphics[width=\columnwidth]{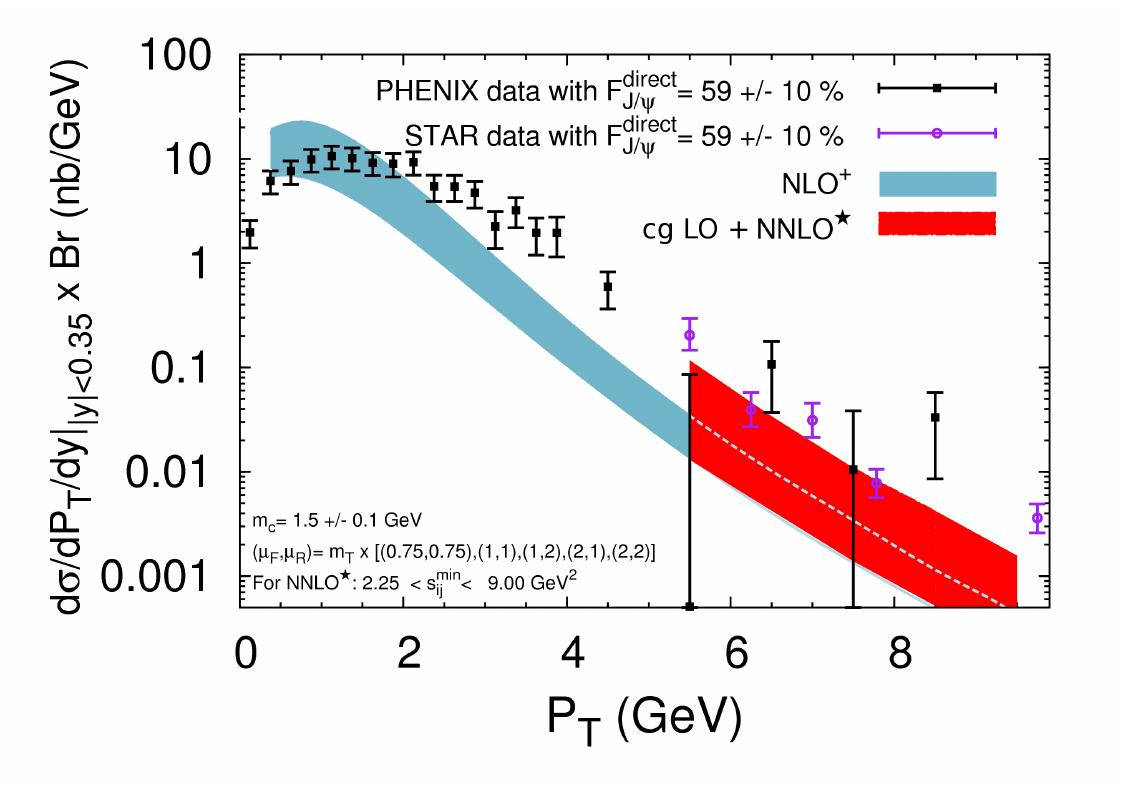}}
\subfloat[forward]{\includegraphics[width=\columnwidth]{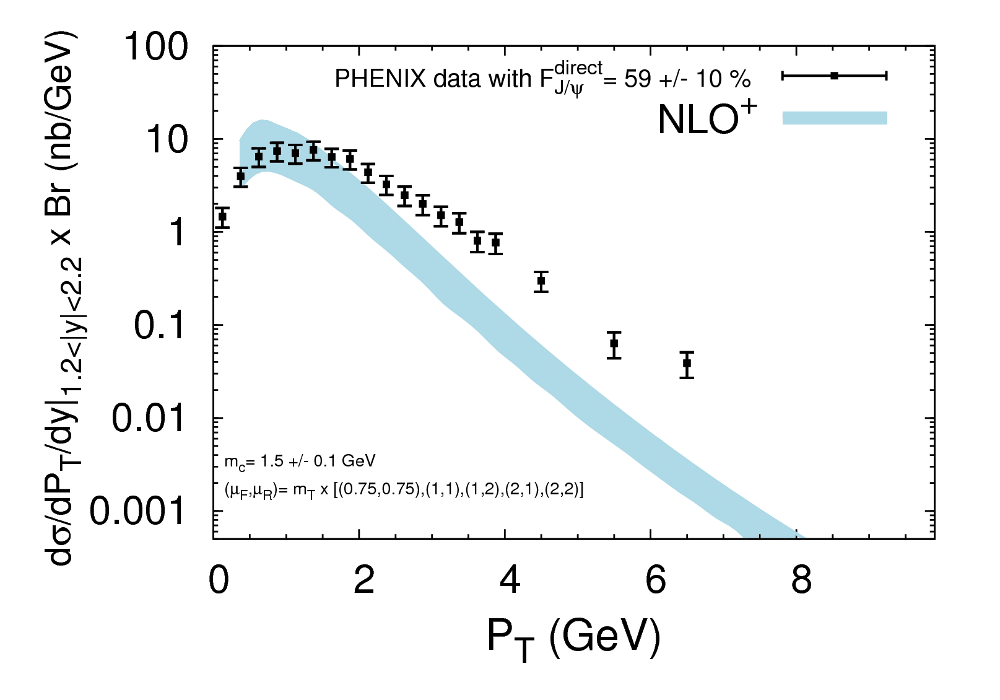}}
\end{center}
\caption{ $d\sigma/dP_T/dy \times {\rm Br}$  in \pp\ collisions
at $\sqrtsNN=200\mathrm{~GeV}$ a) in the central ($|y|<0.35$) and 
b) in the forward ($1.2 < |y|< 2.2$) regions at NLO$^+$ (and $cg$ LO + NNLO$^{\star}$ for (a)) compared to
the PHENIX~\cite{Adare:2006kf} and STAR~\cite{Abelev:2009qa} data. 
The theoretical-error bands come from combining the
uncertainties resulting from the choice of $\mu_f$, $\mu_r$, $m_q$, see text.}
\label{fig:dsigdpt-NLO_plus}
\end{figure*}

\begin{figure*}[htb!]
\begin{center}
\subfloat[central]{\includegraphics[width=\columnwidth]{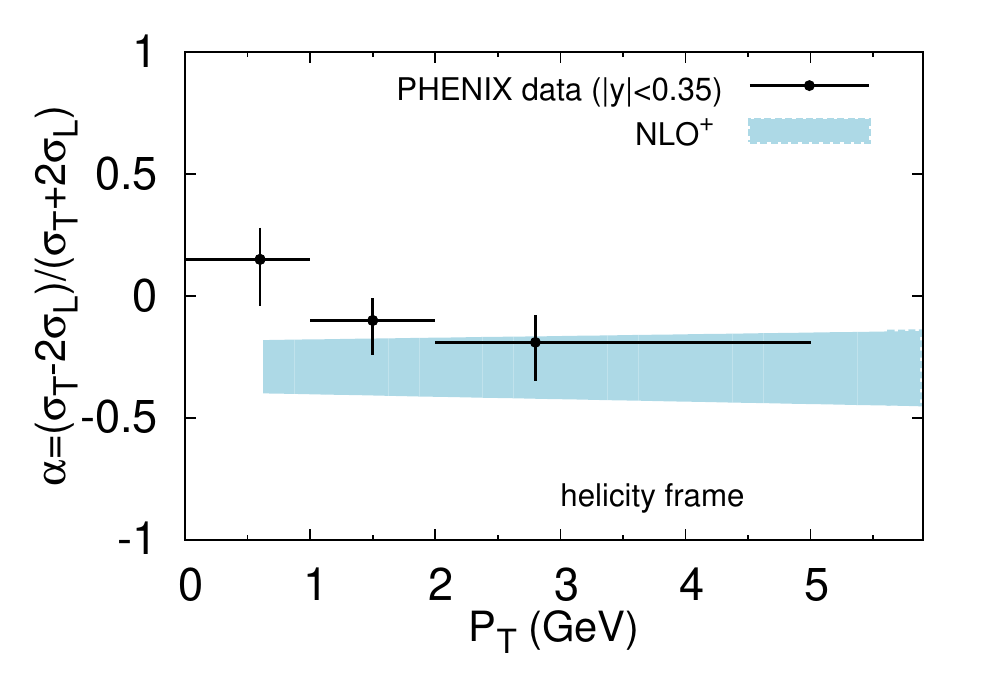}}
\subfloat[forward]{\includegraphics[width=\columnwidth]{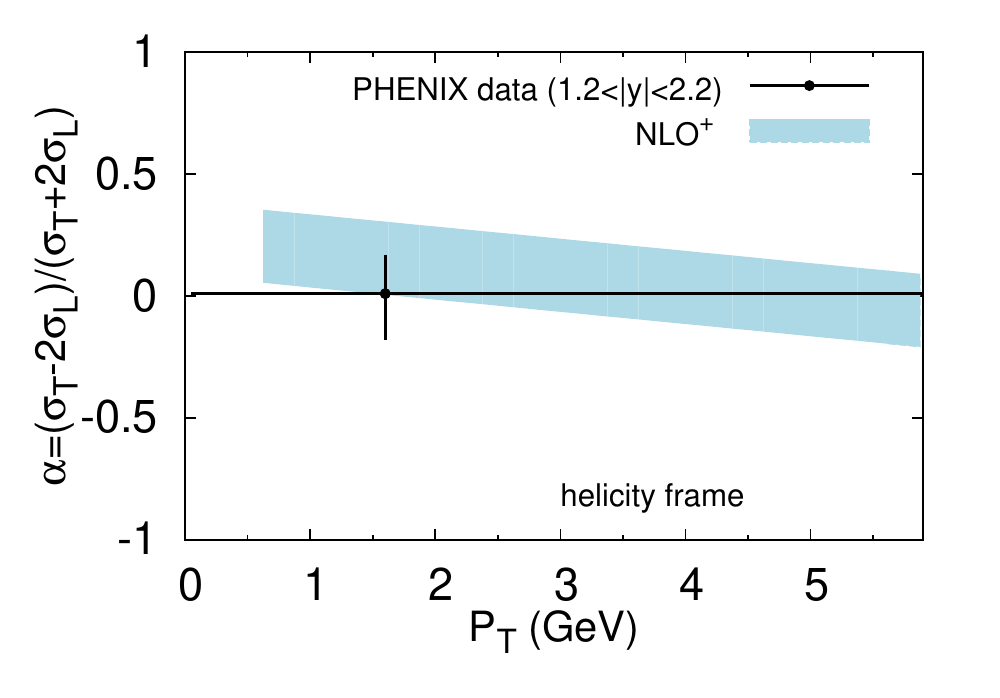}}
\end{center}
\caption{Comparison between $\alpha(P_T)$ for $J/\psi$ directly produced at NLO$^+$ (see text) and the PHENIX 
data~\cite{Adare:2009js,Atomssa:2008dn} in both rapidity regions in \pp\ at $\sqrtsNN=200\mathrm{~GeV}$.}
\label{fig:pol-NLO_plus}
\end{figure*}

In order to obtain the polarisation of the direct yield, one has to combine the polarisation from both
contributions taking into account their proper weight. Schematically one has:
\eqs{ \alpha^{\rm  NLO^+}_{\rm direct}(y,P_T)= \frac{\Delta\sigma^{\rm NLO} \alpha^{\rm NLO} + \Delta\sigma^{cg} \alpha^{cg}}{\Delta\sigma^{\rm NLO}+\Delta\sigma^{cg}}, 
}
where $\Delta \sigma$ is the differential cross section for one given process integrated in a given bin in $y$ and $P_T$.

Beforehand, we present results for the NLO$^+$ yield, namely the yield at NLO accuracy from $gg$ and $gq$ fusion added to the yield
from $cg$ fusion at LO accuracy\footnote{The NLO corrections to $cg \to J/\psi X$ are not yet known.}. The sum of both contributions
differential in $P_T$ is compared to the PHENIX and STAR data on \cf{fig:dsigdpt-NLO_plus} a) and b). In the central region, 
the yield at NNLO$^\star$ from $gg$ and $gq$ fusion is also shown (with $cg$ at LO added). The computation being close to the data for $P_T< 4$ GeV, it is reasonable
to compare them to the PHENIX measurements~\cite{Adare:2009js,Atomssa:2008dn} as done on~\cf{fig:pol-NLO_plus}.

Except for the lowest $P_T$ point in the central region,
the direct NLO$^+$ yield polarisation is compatible with the (prompt) data\footnote{Note also that for 
the $P_T$ bin from 2 to 4 GeV, 
the yield is not perfectly described by the NLO$^+$ -- the deviation is slightly larger than 1 $\sigma$. 
Further contributions, such as the $\alpha_S^5$ ones, may need to be taken into account, 
in turn altering the results presented here for the polarisation.}, indicating a small impact of the $\psi(2S)$ and 
$\chi_c$ feed-downs on the polarisation. The same conclusion holds 
from the $s$-channel cut analysis~\cite{Haberzettl:2007kj,Lansberg:2005pc}, where a good agreement 
with experimental data was also obtained, 
at least in the central region. It is worth recalling that the latter analysis was done by neglecting 
the usual contribution from the CSM which are the purpose of this work and are evidently not negligible. 
Since the results for the polarisation are similar in both analysis, a combined study would 
follow the same trend.

Finally, the disagreement between the data and the present evaluation in the lowest $P_T$ bin 
is not worrisome in view of the expected impact of higher QCD corrections in this bin.

\subsection{Data-driven evaluation of the feed-down effect on the polarisation}

\begin{figure*}[!htb]
\begin{center}
\subfloat[central]{\includegraphics[width=\columnwidth]{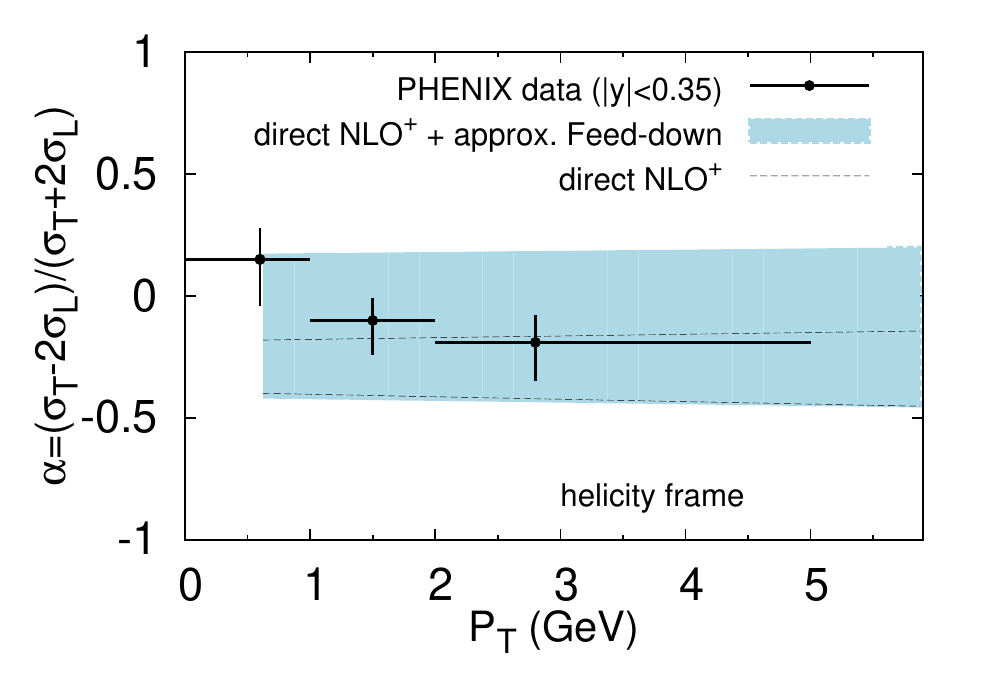}}
\subfloat[forward]{\includegraphics[width=\columnwidth]{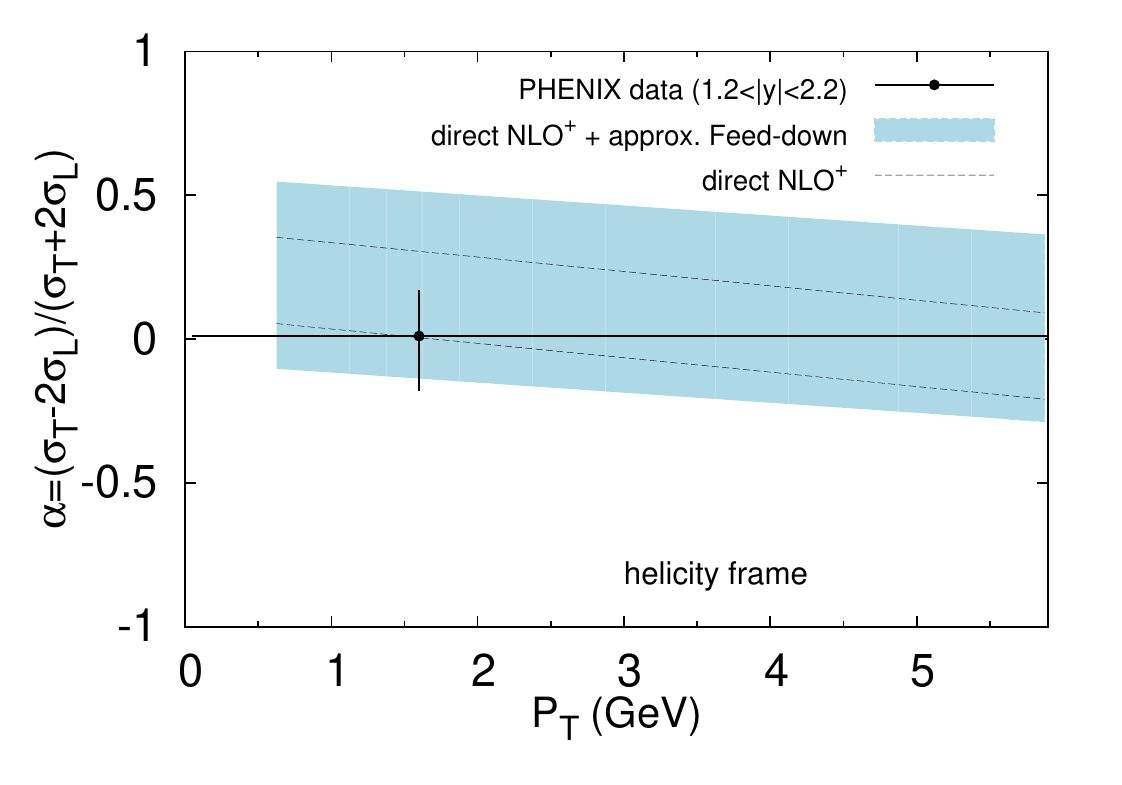}}
\end{center}
\caption{Comparison between the extrapolation of $\alpha$ for prompt $J/\psi$ in \pp\ at $\sqrtsNN=200\mathrm{~GeV}$ (blue band), the range of $\alpha$ for the direct NLO$^+$ (two dashed lines) and the PHENIX measurements in both rapidity regions~\cite{Adare:2009js,Atomssa:2008dn}.}
\label{fig:pol_prompt}
\end{figure*}

For the time being, there exists no measurement of the direct $J/\psi$ polarisation after extraction of both
the $B$ and $\chi_c$ feed-downs. While at RHIC energy the former contributes less than 5 \% of the full yield~\cite{Oda:2008zz}
 and thence should not affect its polarisation, the $\chi_c$ feed-down (up to 30-40  \%) may impact strongly on the observed values
of $\alpha$. It is however possible to constrain its effects by using existing data from Hera-$B$ on 
$\sigma_{\chi_{c1}}/\sigma_{\chi_{c2}}$ 
for instance, relying  on the dominance of the E1 transition between the $\chi_c$'s and the $J/\psi$.

Indeed, using E1 dominance~\cite{Cho:1994gb}, one can obtain a first approximation of the
 yield of longitudinally (transversally) polarised $J/\psi$, denoted  $N^{h=0}_{^3S_1}$ 
(resp. $N^{|h|=1}_{^3S_1}$) from the parent $\chi_c$,  in terms of
 simple\footnote{We should emphasise that these relations neglect effects that could arise
from the difference in the definition of the helicity frame of the $\chi_c$ and that of its 
decay product (the $J/\psi$ here). These effects are expected to vanish for large momenta for the
quarkonium. Such an approximation may not be accurate for the smallest $P_T$'s considered here.}
 relations involving the polarised $\chi_c$ yields, these are 
\eqs{N^{h=0}_{^3S_1}=  &\frac{1}{2}\sigma_{~^3P_1}^{|h|=1} \Br(^3P_1\to \!^3S_1 \gamma) +
\frac{2}{3} \sigma_{^3P_2}^{h=0} \Br(^3P_2\to \!^3S_1 \gamma) \nn \\+
&\frac{1}{2}\sigma_{^3P_2}^{|h|=1} \Br(^3P_2\to \!^3S_1 \gamma) }
and
\eqs{N^{|h|=1}_{^3S_1}&= \sigma_{~^3P_1}^{h=0} \Br(^3P_1\to \!^3S_1 \gamma) 
+\frac{1}{2} \sigma_{~^3P_1}^{|h|=1} \Br(^3P_1\to \!^3S_1 \gamma) \\
&+\frac{1}{3} \sigma_{^3P_2}^{h=0} \Br(^3P_2\to \!^3S_1 \gamma) 
+\frac{1}{2} \sigma_{^3P_2}^{|h|=1} \Br(^3P_2\to \!^3S_1 \gamma)\\
&+\sigma_{^3P_2}^{|h|=2} \Br(^3P_2\to \!^3S_1 \gamma)\nn .}

Let us analyse now the two otherwise extreme cases $\alpha^{\rm max}_{\rm from\ \chi_c}$ and 
$\alpha^{\rm min}_{\rm from\ \chi_c}$:
\begin{itemize}
\item $\alpha^{\rm max}_{\rm from\ \chi_c}$ is reached when $N^{|h|=1}$ is maximised and $N^{h=0}$ minimised. This happens
  when the $^3P_1$ yield is fully $h=0$  and the $^3P_2$ one is fully $|h|=2$. This indeed gives $N^{h=0}_{^3S_1}=0$ and thus 
$\alpha^{\rm max}_{\rm from\ \chi}=+1$.

\item $\alpha^{\rm min}_{\rm from\ \chi_c}$ is reached when $N^{|h|=1}$ is minimised and $N^{h=0}$ maximised. This happens for a $^3P_1$ yield 
fully $|h|=1$ and the $^3P_2$ one fully $h=0$. This gives:
\end{itemize}
\eqs{\label{eq:extrapol-3}
N^{h=0}_{^3S_1}=&\frac{1}{2}  \sigma_{^3P_1} \Br(~^3P_1\to ~^3S_1 \gamma) \\
+&\frac{2}{3} \sigma_{^3P_2} \Br(~^3P_2\to ~^3S_1 \gamma) \\
N^{|h|=1}_{^3S_1}=& \frac{1}{2}
\sigma_{^3P_1} \Br(~^3P_1\to ~^3S_1 \gamma) \\
+&\frac{1}{3} \sigma_{^3P_2} \Br(~^3P_2\to ~^3S_1 \gamma)}

Recently, the Hera-B~\cite{Abt:2008ed} and CDF~\cite{Abulencia:2007bra} collaborations have measured 
the ratio of 
$R_{12}=\frac{\sigma_{\chi_{c1}} \Br(\chi_{c1}\to J/\psi \gamma)}{ \sigma_{\chi_{c2}} \Br(\chi_{c2}\to J/\psi \gamma)}$ 
which we can therefore use in~\ce{eq:extrapol-3}. It is however worth noticing that both results
are somewhat different, $R_{12}=1.0\pm 0.4$ for Hera-B for $0< P_T< 2$ GeV  at $\sqrt{s}=41.6$ GeV 
and $R_{12}=2.5\pm 0.1$ for 
CDF for $P_T >4.0$ GeV at $\sqrt{s}=1.96$ TeV. Since we focus here on low $P_T$ data, we prefer to opt 
for the Hera-B value,  which gives for the central value
\eqs{\alpha^{min}_{\rm from~\chi{_c}}=&\frac{(\frac{1}{3}+R_{12} \frac{1}{2})-2(\frac{2}{3}+R_{12} \frac{1}{2})}{
(\frac{1}{3}+R_{12} \frac{1}{2})+2(\frac{2}{3}+R_{12} \frac{1}{2})}
%\\=\frac{(2+3)-2(4+3)}{(2+3)+2(4+3)}
=-\frac{9}{19}\simeq-0.47
}

Supposing that about $30\%$ of the $J/\psi$ come from $\chi_c$ decays independent of $P_T$ in the range considered here,
we expect a partial contribution to the polarisation ranging from $0.3 \times (+1) $ to $0.3 \times(-0.47)$. Regarding the
other $70\%$, one can simply multiply the result obtained above for the direct yields (section 3.3)
by $0.7$, since in a good approximation the polarisation of $J/\psi$ from $\psi(2S)$ is expected to be 
identical to the direct one.

\subsection{Comparison with the PHENIX data}
 
Combining the result obtained for the direct yield at NLO$^+$ and the extreme expected polarisations of 
$J/\psi$ from $\chi_c$, one can compare (\cf{fig:pol_prompt}) our results for the CSM prompt $J/\psi$ yield with the 
PHENIX data in the central~\cite{Adare:2009js} and the forward~\cite{Atomssa:2008dn} regions.
Within the present admittedly large theoretical uncertainties, the agreement with the data is good.

Three comments are in order: a) the result should not be straightforwardly extrapolated to $P_T=0$,  b) at larger $P_T$, 
$\alpha_S^5$ (for $gg+gq$) and $\alpha_S^4$ (for $cg$) will eventually dominate due to their $P_T^{-4}$ scaling. While 
the former is expected from similar studies~\cite{Artoisenet:2008fc,Artoisenet:2008zza,Lansberg:2008gk} to be longitudinal, 
we have qualitatively checked that the yield from the latter (more precisely from $cg$ fusion at NLO$^\star$) 
is similar to $cg\to J/\psi c$ at LO; c) for BHPS $c(x)$ (showing a peak of the charm distribution around $x\approx 0.1-0.3$), 
one expects a relative enhancement of the yield from  $cg\to J/\psi X$ 
for increasing $P_T$, \ie~for a momentum fraction of the charm quark in the proton approaching the value
where $c^{\rm BHPS}(x)$ peaks. 
Yet, a dedicated study of these effects is needed to draw more quantitative conclusions. 
Indeed, depending on the relative yield of  $gg$ \& $gq$ fusion at NNLO$^\star$ and $cg$ fusion at LO (or NLO$^\star$)
at large $P_T$, the polarisation may end up to be strongly longitudinal or slightly transversal. In any case,  as 
opposed to non-relativistic QCD analyses of including colour octets~\cite{Braaten:1999qk}  (see~\cite{Chung:2009xr} 
for a recent application for RHIC energies)
transitions, the CSM yield is not expected to become strongly transverse for increasing $P_T$.

Our results for $gg$ and $qg$ fusion are in qualitative agreement
 with the analysis carried out by Baranov and Szczurek~\cite{Baranov:2007dw} using the $k_t$
factorisation approach. However the latter study predicts a dominance of $\chi_{c2}$ feed down over the prompt
$J/\psi$ yield which may conflict both with  the preliminary upper value~\cite{Oda:2008zz} for the $\chi_c$  feed down ($<42\%$ at 90\% C.L.) by the PHENIX collaboration
and with the Hera-B results for the ratio $\sigma_{\chi_{c1}}/\sigma_{\chi_{c2}}$.
We further note that some contributions appearing at $\alpha_S^5$ and enhanced by $\log(s)$~\cite{Khoze:2004eu} also produce a yield
which is mainly longitudinal.

%%%%%%%%%%%%%%%%%%%%%%%%%%%%%%%%%%%%%%%%%%%%

\section{Conclusions and outlooks}

In conclusion, we have carried out the first NLO analysis in the 
Colour-Singlet Model of the polarisation  of the direct $J/\psi$ yield at RHIC including contributions from  
$c$-quark--gluon fusion. Our result for the yield differential in $P_T$ is in near 
agreement with the measurement at low and mid $P_T$ 
both in the central and forward rapidity regions. As regards the polarisation, our evaluation
for the direct yield is in good agreement with the PHENIX data for {\it prompt} 
$J/\psi$.  This therefore
points at a small impact of the feed-downs on this observable in this kinematical region.
 
Using constraints from existing data, namely
the fraction of $\chi_c$ feed-down and the ratio $\sigma_{\chi_{c1}}/\sigma_{\chi_{c2}}$, we have determined 
likely extreme values of the polarisation of the $J/\psi$ from $\chi_c$ by relying 
on the E1 dominance of the transition $\chi_c \to J/\psi \gamma$. This has enabled us to extrapolate our 
polarisation evaluation 
of the direct yield to the prompt one. The results obtained are also in good agreement with the PHENIX data.

Motivated by this agreement, the next step would be to study the Cold Nuclear Matter (CNM) effects on 
the yield polarisation in the CSM framework. These effects are expected to come mainly from 
the shadowing of parton distributions and the final-state interactions 
between the $c\bar c$ pair and the CNM. We could for instance take benefit of the Glauber framework 
{\small \sf JIN}~\cite{Ferreiro:2008qj} used  in \cite{Ferreiro:2008wc} to study the CNM effects on the $J/\psi$ yield in
$dA$ and $AA$ collisions at RHIC; the latter can indeed deal with the polarisation information. 
Such a study would surely be very expediently done
in order to refine the arguments presented in~\cite{Ioffe:2003rd} to employ quarkonium polarisation in 
heavy-ion collisions as a possible signature of the quark-gluon plasma.

%%%%%%%%%%%%%%%%%%%%%%%%%%%%%%%%%%%%%%%%%%%%

\section*{Acknowledgments}

 We thank  S.J. Brodsky, Z. Conesa del Valle, P. Faccioli, F. Fleuret, R. Granier de Cassagnac, A.Kraan,
J. Lee, C. Louren\c co, A. Linden-Levy, B.Pire, L.~Szymanowski for useful discussions,
J.~Campbell, F.~Maltoni and F.~Tramontano for their NLO code, as well as
J. Alwall and P. Artoisenet for useful technical advice.

%%%%%%%%%%%%%%%%%%%%%%%%%%%%%%%%%%%%%%%%%%%%

%%%%%%%%%%%%%%%%%%%%%%%%%%%%%%%%%%%%%%%%%%%%%%%%
%% BACKMATTER
%%%%%%%%%%%%%%%%%%%%%%%%%%%%%%%%%%%%%%%%%%%%%%%%

\end{document}